\documentclass[preprint,preprintnumbers,showpacs,amsmath,amssymb]{revtex4-1}

\usepackage{graphicx}
\usepackage{dcolumn}
\usepackage{bm}

\begin{document}


\title{Inelastic cross sections, overlap functions and $C_{q}$ moments from ISR to LHC energies in proton interactions}


\author{P.C. Beggio$^{}$}
\affiliation{ $^{}$Laborat\'orio de Ci\^encias Matem\'aticas -
LCMAT, Universidade Estadual do Norte
Fluminense Darcy Ribeiro - UENF, 28013-602, Campos dos Goytacazes, RJ, Brazil \\
}


\begin{abstract}

We investigated the energy dependence of the parton-parton inelastic
cross sections, parton-parton inelastic overlap functions and the
$C_{q}$ moments in proton interactions from $\sqrt{s}$= 10 to 14000
$GeV$. The used approach is based on a phenomenological procedure
where elastic and inelastic proton observables are described in a
connected way by exploring the unitarity of $S$-Matrix. Applying a
Quantum Chromodynamics inspired eikonal model, that contains
contributions of the quark-quark, quark-gluon and gluon-gluon
interactions, theoretical predictions on inelastic cross sections
and $C_{q}$ moments are compared with measurements showing
successfully description of the experimental data. The KNO
hypothesis violation is discussed as a consequence of the semihard
contribution to the multiparticle production in the interactions, in
accordance to several experimental and theoretical previous results.
Prediction to the ratio $\sigma_{el}/\sigma_{tot}$ as a function of
the collision energy is presented and also compared with the
experimental information.

\end{abstract}



\maketitle

\section{Introduction}

In the nonperturbative sector of Quantum Chromodynamics - QCD - one
of the problems is the hadronization mechanism and the probability
for producing the number $n$ of charged hadrons in the final state
of proton collisions, $P_{n}$, is a very important physical
observable to investigate the multiparticle production dynamics
sector, providing important insigths on the particle production
mechanisms. The multiplicity distribution is defined as
$P_{n}=\sigma_{n}/\sigma_{in}$, where $\sigma_{n}$ and $\sigma_{in}$
are the topological and inelastic cross sections, respectively. The
cross sections, and hence $P_{n}$, cannot yet be calculated from
QCD. Thus, our knowledge on multiparticle production dynamics is
still phenomenological and based on a wide class of models
\cite{Fiete} and some theoretical principles. In particular the
unitarity principle is very important in the nonperturbative sector,
which regulates the relative strength of elastic and inelastic
processes \cite{Troshin}. We note that the Froissart-Martin bound on
the high energy behavior of total cross section, $\sigma_{tot}$,
first derived by Froissart from the Mandelstam representation
\cite{Foissart61} and then proved directly by Martin using unitarity
and analycity \cite{Martin66}, has been extended to the
$\sigma_{in}$, also implying that the $\sigma_{in}$ cannot rise
faster than $ln^{2}(s)$ \cite{AMartin}.

From experimental side, LHC data on both $\sigma_{in}$ and charged
particle multiplicity moments, in restrict pseudorapidity intervals,
are available by CMS, ATLAS, ALICE, TOTEM and LHCb experiments
\cite{CMS2011} \cite{Aad001} \cite{Abelev001}  \cite{Antchev001}
\cite{Antchev002} \cite{Antchev003} \cite{CMS2013} \cite{LHCb}. In
addition, there are experimental results on these quantities at
lower center-of-mass energy, $\sqrt{s}$, in both restrict and full
phase space pseudorapidity intervals obtained by ABCDWH, UA5 and
E735 Collaborations \cite{Amaldi} \cite{ABC} \cite{UA5kno546}
\cite{UA5} \cite{UA5kno200} \cite{Alexopoulos} \cite{barroso}.
Nowdays, the understanding of the rise of $\sigma_{in}$ with
$\sqrt{s}$, as well as $P_{n}$, are of fundamental importance for
hadron collider physics, particle astrophysics \cite{Aad001} and
also for multiparticle production dynamics sector of QCD
\cite{Erasmo001} \cite{GUNPB}. For these reasons, it is important to
look for approaches and to test calculation schemes able to
describe, in an unified way, elastic and inelastic physical
observables in the wide interval of $\sqrt{s}$ covered by
experiments. It can give us the chance to quantify the most
important trends of the experimental points, as well as indicating
for new possible theoretical developments.

In the present work we applied a phenomenological procedure where
the connection between elastic and inelastic channels is established
through the unitarity condition using the eikonal approximation
\cite{lam01} \cite{BeggioMV}. We have adopted a QCD-inspired eikonal
function \cite{BeggioLuna} in order to investigate the approximate
collision energy dependence of the quark-quark, quark-gluon and
gluon-gluon of both, inelastic overlap functions and inelastic cross
sections. In the present analysis all parameters of the eikonal has been determined carrying
out a global fit to all high energy forward $pp$ and $p\overline{p}$
scattering data above $\sqrt{s}$=10 $GeV$ \cite{BeggioLuna}. We give
also quantitative results on the $\sqrt{s}$ dependence of the
$C_{q}$ moments of $P_{n}$. The mentioned phenomenological procedure
is here referred to as Geometrical Approach and it has permitted
some previous studies involving the hadronic physics \cite{BeggioMV}
\cite{BeggioHama} \cite{BeggioBPJ2008} \cite{BeggioNPA2011}
\cite{BeggioNPA2013}.

The paper is organized as follows: in the next section we present
the main equations of the Geometrical Approach, which is composed by
the two models, the QCD motivated eikonal dynamical gluon mass model
(elastic channel) and the string model (inelastic channel).
Motivated by the previous good results obtained by using the
approach \cite{BeggioLuna}, in Section III we present the strategy
used to calculate the partonic and total inelastic overlap function
as well as the partonic and total inelastic cross sections. In the
sequence we calculate the $Cq$ moments of the multiplicity
distributions discussing our main results. In Section IV we draw our
conclusions.

\section{Geometrical Approach}

The approach treats the protons as composite and extended objects
and, hence, the impact parameter $b$ is used as an essential
variable in the description of the collisions \cite{BeggioMV}
\cite{BeggioHama} \cite{BeggioBPJ2008}, where the impact parameter
denotes the distance between the centers of the colliding composite
systems in the plane perpendicular to the beam direction
\cite{barshayrein}. For the sake of discussions we presented here
the main equations of the approach, discussed in details in
\cite{BeggioLuna}. The $\chi_{I}(s,b)$ represents the imaginary
eikonal and it has been obtained from a QCD-inspired eikonal model
which is completely determined only by elastic fit and incorporates
soft and semihard process using a formulating compatible with
analyticity and unitarity principles. This eikonal model is referred
to as Dynamical Gluon Mass (DGM) model \cite{luna008} \cite{luna006}
\cite{luna009} and it has been written in terms of even and odd
eikonal parts, connected by crossing symmetry, and this combination
reads
\begin{equation}
\chi_{pp}^{\overline{p}p}{(s,b)}=\chi^{+}{(s,b)}\pm\chi^{-}{(s,b)}\,.
\end{equation}
The odd eikonal is written to account on the difference between $pp$
and $p\overline{p}$ channels at low energies and it is simply
\begin{equation}
\chi^{-}{(s,b)}=C^{-}\sum\frac{m_{g}}{\sqrt{s}}e^{i\pi/4}W(b;\mu^{-}).
 \label{sigma1}
\end{equation}
$W(b;\mu_{ij})$ is the overlap density for the partons at impact
parameter $b$, $(i,j=q,g)$ and $m_{g}=364$ $MeV$ is an infrared
gluon scale mass \cite{luna010}. In Eq. (1) the even eikonal is
written as the sum of quark-quark, quark-gluon and gluon-gluon
contributions
\begin{equation}
\chi^{+}{(s,b)}=\chi_{qq}{(s,b)}+\chi_{qg}{(s,b)}+\chi_{gg}{(s,b)}\,.
\end{equation}
The functions $\chi_{qq}{(s,b)}$ and $\chi_{qg}{(s,b)}$ are needed
to describe the lower energy experimental points and, based on the
Regge phenomenology, it has been parametrized as
\begin{equation}
\chi_{qq}  (s,b) = i \Sigma \, A \, \frac{m_{g}}{\sqrt{s}}
W(b;\mu_{qq}),
\end{equation}
and
\begin{equation}
\chi_{qg}  (s,b) = i \Sigma \left[ A' + B' \, \ln \left(
\frac{s}{m_{g}^{2}} \right) \right] \, W(b;\mu_{qg}).
\end{equation}
The $\Sigma$ factor is defined as
$\Sigma=9\pi\bar{\alpha}_{s}^{2}(0)/m_{g}^{2}$, being
$\bar{\alpha}_{s}$ and $m_{g}$ non-perturbative quantities. At
higher energies the perturbative component of the DGM model is
dominated by gluons with a very small fractional momentum and it is
given by
\begin{equation}
\chi_{gg}(s,b) = \sigma_{gg}(s) W(b;\mu_{gg}) \,.
\end{equation}
From the last equation we can see that the energy dependence of
$\chi_{gg}(s,b)$ comes from the gluon-gluon cross section,
$\sigma_{gg}(s)$, which gives the main contribution to the
asymptotic behavior of parton-parton total cross sections. In DGM
model formulation the gluon-gluon cross section has been obtained
from QCD parton model perturbative cross section for parton pair
colliding, also used in some previous works \cite{varias}
\cite{Margolisviola} \cite{BlockNPb} \cite{BlockGregores}. Thus,
defining the variable $\tau = x_{1}x_{2} =\hat{s}/s$ we can obtain
\begin{eqnarray}
\sigma_{gg}(s) = \int_{0}^{1} d\tau \, \left[
\int\,\int\,g(x_{1})\,g(x_{2})\,\delta(\tau-x_{1}x_{2})\,dx_{1}dx_{2}
\right]\,\hat{\sigma}_{gg}(\hat{s})\,\Theta(\hat{s}-m_{g}^{2}),
\end{eqnarray}
where $x_{i}$ is the fraction of the momentum of the proton carried
by gluon $i$, the factor $\hat{\sigma}_{gg}(\hat{s})$ is the total
cross section for the subprocess $gg\rightarrow gg$ calculated
through the dynamical perturbation theory \cite{Pagels}
\cite{luna008} \cite{luna006}, while the Heaviside function
determines a threshold to gluon mass production. In turn, the
integral in brackets is the convoluted structure function for pair
gluon-gluon, where using $x_{2}=\tau/x_{1}$ we can write
\begin{eqnarray}
F_{gg}(\tau )\equiv [g\otimes g](\tau)=\int_{\tau}^{1}
\frac{dx}{x}\, g(x)\, g\left( \frac{\tau}{x}\right).
\end{eqnarray}
We recall that $F_{gg}(\tau )$ counts the number of gluons in the
colliding protons \cite{BlockNPb}. With respect the phenomenological
gluon distribution we adopted the form
\begin{eqnarray}
g(x)= N_{g}\,\frac{(1-x)^{5}}{x^{J}}\,,
\end{eqnarray}
with $N_{g}=\frac{1}{240}(6-\epsilon)(5-\epsilon)...(1-\epsilon)$
and $J=1+\epsilon$=1.21. It is interesting to note that, due to the
relation $x\simeq m_{g}/\sqrt{s}$, the $\sigma_{gg}$ all energy
dependence comes from the small $x$ behavior of $g(x)\sim x^{-J}$,
which becomes increasing important as energy increases. Hence,
higher $\sqrt{s}$ means smaller $x$ and therefore gluons with
smaller momentum fractions are more abundant and this is the origin
of the rising cross section in this approach \cite{BlockNPb}. Thus
the gluon-gluon cross section represents the probability for the
subprocess $gg\rightarrow gg$ when protons are colliding at each
other.

The DGM model parameters in Eqs. (1) - (9) has been obtained in
\cite{BeggioLuna} carrying out a global fit to all forward $pp$ and
$p\overline{p}$ scattering data above $\sqrt{s}$= 10 $GeV$, namely
total cross section $\sigma_{tot}^{pp,p\overline{p}}$, the ratio of
the real to imaginary part of the forward scattering amplitude
$\rho^{pp,p\overline{p}}$, the elastic differential scattering cross
sections $d\sigma^{p\overline{p}}/dt$ at $\sqrt{s}= 546$ $GeV$ and
$\sqrt{s}=1.8$ $TeV$ as well as the TOTEM datum on
$\sigma_{tot}^{pp}$ at $\sqrt{s}=7$ $TeV$ \cite{totem}, where we set
the value of the gluon scale mass to $m_{g}$= 364 $MeV$
\cite{luna010} and fixed also the values of $n_{f}$= 4 and
$\Lambda$= 284 $MeV$. All fitted parameters are reproduced from
\cite{BeggioLuna} in Table I, but now including the values of the
mentioned fixed quantities ($m_{g}$, $\Lambda$, $J$ and
$\hat{\alpha}_{s}(0)$). The $\chi^{2}/DOF$ for the global fit was
0.98 for 320 degrees of freedom. We would like to emphasize that we
have proceeded the analysis in \cite{BeggioLuna} in order to
consider the TOTEM datum on $\sigma_{tot}^{pp}$ at $\sqrt{s}=7$
$TeV$, hence the parameters obtained are different from those at
\cite{luna008}.
\begin{table*}
\caption{Values of the DGM model parameters from the global fit to
the scattering $pp$ and $\bar{p}p$ reproduced from
\cite{BeggioLuna}. The result was obtained by fixing the values of
$m_{g}$, $\Lambda$, $J$ and $\hat{\alpha}_{s}(0)$.}
\begin{ruledtabular}
\begin{tabular}{cc}
$C_{gg}$ & (1.62$\pm$0.37)$\times 10^{-3}$ \\
$\mu_{gg}$ [GeV]& 0.642$\pm$0.034 \\
$A$ & 9.04$\pm$4.94 \\
$\mu_{qq}$ [GeV] & 1.299$\pm$0.797 \\
$A^{\prime}$ & (4.68$\pm$1.89)$\times 10^{-1}$ \\
$B^{\prime}$ & (4.53$\pm$1.94)$\times 10^{-2}$ \\
$\mu_{qg}$ [GeV] & 0.825 $\pm$0.015 \\
$C^{-}$ & 3.12$\pm$0.33 \\
$\mu^{-}$ [GeV]& 0.799$\pm$0.298 \\
\hline
$m_{g}$ [MeV]& 364 \\
$\Lambda$ [MeV]& 284 \\
$J$ & 1.21 \\
$\hat{\alpha}_{s}(0)$ & 0.801 \\
\hline
$\chi^{2}/DOF$  &  0.98 \\
\end{tabular}
\end{ruledtabular}
\end{table*}

In order to study multiplicities we recall that the shape of $P_{n}$
is so complicated that is difficult to get any analytical expression
for it from solution of QCD equations \cite{DreminAN}. However, an
alternative approach is possible by studies of moments of
distribution, since all moments together contain the information of
the full distribution \cite{Fiete} and it facilitates the comparison
with other models, allowing also studies of the KNO scaling
hypothesis which has been an important phenomenological issue on the
energy dependence of the $P_{n}$. However, a subtle problem is the
proper choice of the set of moments to be adopted in the analysis
(for an instructive discussion on the moments of a distribution see
\cite{Fialkowski}). In this work we have adopted the standart choice
and used the simple power moments, defined by
\begin{equation}
<n^{q}>\,=\,\sum_{n}\, n^{q}\,P_{n}\, ,
\end{equation}
and their scaled version
\begin{equation}
C_{q}\,=\,\frac{<n^{q}>}{<n>^{q}}\,=\,\frac{\sum\,
n^{q}\,P_{n}}{[\sum\,n\,P_{n}]^{q}}
\end{equation}
to parametrize $P_{n}$. Here $q$ is a positive integer and referred
to as the rank or order of the moment. The use of the average
multiplicity and a few lowest $C_{q}$ moments ($q=$ 2,..,5) allows
to parametrize the $P_{n}$ quite satisfactorily \cite{Fialkowski}.
Quantitative predictions for $H_{q}$ moments oscillations were
presented in \cite{BeggioLuna} \cite{BeggioNPA2013} applying the
Geometrical Approach. Now, to calculate the $C_{q}$ moments, Eq.
(11), we have applied the so-called string model which enables
linkage between the $P_{n}$ and $\chi_{I}(s,b)$ \cite{BeggioMV}
\cite{BeggioLuna}. In this model $P_{n}$ is decomposed into
contributions from each $b$ with weight $1-e^{-2\,\chi_{I}(s,b)}$
and written as
\begin{equation}
P_{n}(s)={\frac
  {\int d^2b\,{\frac{[1-e^{-2\,\chi_{I}(s,b)}]}{f(s,b)}}\,
    \phi^{(1)}\!\left(\frac{z}{f(s,b)}\right)}
  {<n(s)>\,\int d^2b\,[1-e^{-2\,\chi_{I}(s,b)}]}},
 \label{Phi}
\end{equation}
where
\begin{equation}
f(s,b)=\xi(s)\,[\chi_{I}(s,b)]^{2A}\,,
\end{equation}
\begin{equation}
 \xi(s)=\frac
  {\int d^2 b\,[1-e^{-2\,\chi_{I}(s,b)}]}
  {\int d^2 b\,[1-e^{-2\,\chi_{I}(s,b)}]\,[\chi_{I}(s,b)]^{2A}}\,,\
 \label{xi}
\end{equation}
and
\begin{equation}
 \phi^{(1)}(z)=2\,\frac{k^k}{\Gamma(k)}\left[\frac{z}{f(s,b)}\right]^{k-1}e^{-k[\frac{z}{f(s,b)}]}\,.
 \label{psi1}
\end{equation}
$<n(s)>$ is the hadronic average multiplicity computed from
experimental values using the Eq. (10), $z=n/<n(s)>$ represents the
usual KNO scaling variable. The function $ \xi(s)$, Eq. (14),
results from normalization condition on $P_{n}$ \cite{BeggioMV}, and
both $k$ and $A$ are fitted parameters discussed in
\cite{BeggioLuna}. The Eq. (15) corresponds to the KNO form of the
negative binomial distribution \cite{Fiete} and $\Gamma$ is the
usual gamma function. The choose of $\phi^{(1)}$, characterized by
$k$ parameter, is motivated by the fact that this distribution
arises as the dominant part of the solution of the QCD equation for
three gluon branching process in the very large $n$ limit
\cite{durand001}, allowing yet a connection between the geometrical
phenomenological approach and the underlying theory of parton
branching \cite{barshay01}.

As mentioned before the string model enables linkage between $P_{n}$
and $\chi_{I}$ and, as physical scenario, asserts that $P_{n}$ in
full phase space can be constructed by summing contributions from
parton-parton collisions occurring at $b$ and $\sqrt{s}$. The
semihard partons produced at the interaction point fly away from
each other a yielding color string, per collision, which breaks up
producing the observed hadrons.

\section{Results and discussion}

\subsection{Inelastic overlap functions}

From Eqs. (1) - (3) we see that the eikonal function is written as
the sum of the components, explicitly
\begin{equation}
\chi_{pp}{(s,b)}=\chi_{qq}{(s,b)}+\chi_{qg}{(s,b)}+\chi_{gg}{(s,b)}
-\chi^{-}{(s,b)}\,.
\end{equation}
In the eikonal representation the inelastic overlap function,
$G_{in}(s,b)$, is related to the imaginary eikonal
\begin{equation}
G_{in}(s,b)=1-e^{-2\,\chi_{_{I}}(s,b)}\equiv \sigma_{in}(s,b)\,
\end{equation}
and it represents the probability of absorption associated to each
$b$ and $\sqrt{s}$. In view of the Eq. (16) we can express the last
equation in terms of the parton components as
\begin{equation}
G_{in}(s,b)=1-e^{-2\,[\,\chi_{_{I}\,;\,qq}(s,b)\,+\,\chi_{_{I}\,;\,qg}(s,b)\,+\,\chi_{_{I}\,;\,gg}(s,b)\,-\chi_{I}^{-}{(s,b)}]}.
\end{equation}
The odd eikonal is proportional to $\frac{1}{\sqrt{s}}$, Eq. (2),
and at ISR energies it represents only about 1 percent in the value
of $\sigma_{in}$ and therefore, as first approximation, we assume
$\chi^{-}= 0$ in this analysis. Thus, in order to investigate the
parton-parton inelastic overlap functions dependence on the
$\sqrt{s}$, a phenomenological procedure is possible by fixing a
component of interaction and maintaining the others equal to zero.
More specifically and as example, to compute $G_{in\,;\,gg}$ we have
\begin{equation}
G_{in\,;\,gg}(s,b) \approx 1-e^{-2\,
\chi_{_{I}\,;\,gg}(s,b)\,}\equiv \sigma_{in \,;\,gg}\,(s,b)\,,
\end{equation}
with $\chi_{_{I}\,;\,qq}=\chi_{_{I}\,;\,qg}=0$ in the Eq. (18). In a
general notation
\begin{equation}
G_{in\,;\,ij}(s,b)\approx 1-e^{-2\,\chi_{_{I}\,;\,ij}(s,b)\,}\equiv
\sigma_{in \,;\,ij}\,(s,b)\,,
\end{equation}
with $(i,j)=(q,g)$. At this point we would like to draw attention
that the Eq. (20) does not represents the partonic probability of
absorption associated to each $b$ and $\sqrt{s}$, in strict analogy
to the Eq. (17). However, with this mathematical procedure we may to
interpret the Eq. (20) as an associated probability with inelastic
events due to the quark-quark or quark-gluon or gluon-gluon
interactions taking place at $b$ and $\sqrt{s}$, which may give
important insights on partonic behavior as function of the
$\sqrt{s}$ in proton interactions. Thus, within this procedure we
are able to study the partonic inelastic overlap functions
dependence on the collision energy, as well as the partonic
inelastic cross sections.

With the QCD-inspired DGM eikonal function, Eqs. (1) - (9), and
fixing the value of $b \sim 0$ we have computed $G_{in\,;\,qq}$,
$G_{in\,;\,qg}$ and $G_{in\,;\,gg}$ applying the Eq. (20) and
computed also $G_{in}$ given by Eq. (18). Fig. 1 shows the results
as function of $\sqrt{s}$. The probability for inelastic events by
quark-quark interactions has appreciable chance of occurrence from
$\sqrt{s}$ = 10 to $\sim$ 500 $GeV$. At ISR this probability varies
from 25 to 45 percent approximately, while the probability of
gluon-gluon interactions is less then 10 percent at the same
interval. Interestingly, above 100 $GeV$ the probability for
inelastic events due to gluon-gluon interactions grows rapidly with
$\sqrt{s}$. From $\sqrt{s}$ = 100 to 500 $GeV$ this probability
varies from $\approx$ 17 to more than 40 percent. At the LHC
energies above 8000 $GeV$ the probability of interactions between
gluons is more than 90 percent. In turn, the quark-gluon inelastic
overlap function behavior seems to indicate a slow logarithimic
growth of the quark-gluon interaction activity at interval of
$\sqrt{s}$ studied. The term $ln(s/m_{g}^{2})$ in Eq. (5) can be
explained by the presence of a massive gluon in the $qg \rightarrow
qg$ subprocess \cite{luna008}. We call attention that parton-parton
scattering processes containing at least one gluon in the initial
state are important to compute the QCD cross sections. In this
respect and as discussed in \cite{lunaBahiaBroilo}, the gluon-gluon
($gg \rightarrow gg$) and quark-gluon ($qg \rightarrow qg$)
scattering in fact dominate at high energies.

The UA5 Collaboration showed that the KNO scaling law was clearly
broken in $p\overline{p}$ collisions at $\sqrt{s}=$ 546 $GeV$
\cite{UA5kno546} \cite{UA5} and this result was confirmed also at
$\sqrt{s}=$ 200 and 900 $GeV$ \cite{UA5kno200}. Related to KNO
hypothesis the physical scenario that emerges in this analysis is
that the fast growing of $G_{in\,;\,gg}$, above 100 $GeV$, increases
significantly the probability of perturbative small-$x$ gluon-gluon
collisions and it may lead to the appearance of minijets. Hence, the
results seems to demonstrate that the KNO violation is a consequence
of the semihard contribution manifestation in the multiparticle
production mechanisms, as $\sqrt{s}$ is increased. However, we
should emphasize that it is not a new result from this work. In
fact, in the defining the Eq. (9) the term $\sim 1/x^{J}$ simulates
the effect of scaling violations in the small $x$ behavior of $g(x)$
\cite{Margolisviola} \cite{BlockNPb} \cite{BlockGregores}
\cite{luna008}, what is reflected in the present analysis.

It is worth mentioning that QCD inspired models are one of the main
theoretical approaches to explain the observed increase of hadronic
cross sections \cite{giulia001} \cite{Daniel001} \cite{Daniel002}.

\subsection{Inelastic cross sections}

The inelastic cross section due to the imaginary eikonal can be
represented by
\begin{equation}
\sigma_{in}(s)=\int\,d^{2}b\,G_{in}(s,b)=\int\,d^{2}b\,[\,1-e^{-2\,\chi_{_{I}}(s,b)\,}]\,.
\end{equation}
Thus, integrating the Eq. (20) in the $b$ plane, we obtain an
associated parton-parton inelastic cross section
\begin{equation}
\sigma_{in\,;\,ij}(s) \approx \int\,d^{2}b\,G_{in\,;\,ij}(s,b)=\int\,d^{2}b\,[1-e^{-2\,\chi_{_{I}\,;\,ij}(s,b)}]\,,
\end{equation}
which allows studies of the approximate energy dependence of the
partonic cross sections. In fact, by using the $\chi_{_{I}}$ given
by Eqs. (1) - (9) we have obtained theoretical predictions about
$\sigma_{in\,;\,qq}$, $\sigma_{in\,;\,qg}$ and $\sigma_{in\,;\,gg}$
dependence on the $\sqrt{s}$ within this procedure, as displayed in
the Fig. 2. The inelastic cross section due to the interactions
between quarks, $\sigma_{in\,;\,qq}$, decreases as $\sqrt{s}$
increases, at $\approx$ 200 $GeV$ we see that $\sigma_{in\,;\,qq}$
$\rightarrow$ 0, reflecting the $G_{in\,;\,qq}$ behavior. Above
$\approx$ 100 $GeV$ the gluon-gluon inelastic cross section rises
fastly as function of the $\sqrt{s}$. In turn we have applied Eq.
(21), which results from unitarity condition, and computed the total
inelastic cross section as function of $\sqrt{s}$ with $\chi_{_{I}}$
given by Eqs. (1) - (9) and by using the parameter set obtained in
\cite{BeggioLuna}, reproduced in Table I. The result is compared
with several measurements in Fig. 2, which is in very good agreement
with the experimental data, specially at the highest energies where
the $\chi_{gg}(s,b)$ contribution determines the asymptotic behavior
of $\sigma_{in}$. It should be stressed that the $\sigma_{in}$ curve
in Fig. 2 has not been fit to data, and also that the $\sigma_{in}$
dependence on the $\sqrt{s}$ has been determined only from fits to
measurements of elastic channel observables. In addition, we have
compared our results from Eq. (21) to that one from the Ref.
\cite{Erasmo001} at some specific energies represented by triangles
in the Fig. 3. We see that both theoretical predictions agree very
well. These results are very encouraging in order to study the
multiplicity distributions $P_{n}$. However, the $\sigma_{in}$
results obtained in this work applying the DGM one-channel model
deserves some comments. The use of one-channel models is limited and
fail to simultaneously describe the total and the elastic cross
section with the same parameter set \cite{Daniel001}
\cite{Daniel002}, thus multichannel models are needed to describe
the diffractive component of the cross section. In this respect, it
was pointed out in \cite{Daniel002} that good descriptions of all
the components of the cross section has been obtained in
\cite{Koze001} \cite{Lipari001} \cite{Lipari002} and also in
\cite{lunaKMR2009} \cite{lunaKMR2010} through multi-channel
formulations. Now, based on the possible relation between the
Poisson distribution of independent collisions and diffractive
processes a suggestion was made \cite{Daniel001} \cite{Daniel002}
that the integrand in Eq. (21) can be identified with a sum of
totally independent collisions. Specifically
\begin{equation}
\sigma_{in}(s)=\int\,d^{2}b\,[\,1-e^{-2\,\chi_{_{I}}(s,b)\,}]=\int\,d^{2}b\,\left[\sum_{n=1}^{\infty}\frac{(\overline{n}_{(s,b)})^{n}e^{-\overline{n}_{(s,b)}}}{n!}
\right]\,,
\end{equation}
where $\overline{n}(s,b)$ is the average number of collisions and
the authors have interpreted it suggesting that the diffractive or
other quasi-elastic processes might have been excluded from
integration in Eqs. (21)/(23). On the basis of analysis done in this
paper, we argue that de success of the DGM model to simultaneously
describe total and inelastic cross sections, with the same parameter
set, may be an indication that the real part of the eikonal
function, $\chi_{R}$, is in fact needed in order to determine all
the parameters of the eikonal model, which satisfactorily seems to
describe the full inelastic cross section through the entire
available $\sqrt{s}$ interval. Thus, the present result on
$\sigma_{in}$ is a straightforward consequence of the overall
parameter fitting of the total cross section, the ratio of the real
to imaginary part of the forward scattering amplitude and the
differential cross section.

Due to the mentioned limitations on the application of the one
channel approaches \cite{Lipari001} it seems instructive show how
the one-channel DGM model can be used to predict the ratio between
the elastic and total cross sections, $\sigma_{el}/\sigma_{tot}$,
which provides crucial information on the asymptotic properties of
the hadronic interactions \cite{Daniel003}. We have used the
equations given by
\begin{equation}
\sigma_{el}(s)=\int\,d^{2}b\,|\,1-e^{-\chi_{_{I}}(s,b)+\,i\,\chi_{_{R}}(s,b)}|^{2}\,,
\end{equation}
\begin{equation}
\sigma_{tot}(s)=2\,\int\,d^{2}b\,[\,1-e^{-\chi_{_{I}}(s,b)}\,Cos\chi_{R}(s,b)]\,,
\end{equation}
to compute this physical quantity. The results of the computation
are displayed in Fig. (4) and reproduces with a good approximation
the experimental information compiled in \cite{Daniel003}.

\subsection{$C_{q}$ moments}

Some characteristics of $P_{n}$ can be quantified in terms of the
scaled version of the simple power moments, Eq. (11). Thus, we have
calculated both theoretical and experimental $C_{q}$ moments for
full phase space $P_{n}$ over a large range of energies where there
are available experimental data, namely at $\sqrt{s}=30.4, 44.5,
52.6, 62.2, 300, 546, 1000$ and $1800$ $GeV$ \cite{ABC}
\cite{Alexopoulos}. For energies $\sqrt{s}\geq 300$ $GeV$ we used
$P_{n}$ data from the E735 Collaboration since it is statistically
more reliable in the high multiplicity region \cite{Alexopoulos}.
Theoretical $P_{n}$ values were obtained by using the mentioned
string model, Eqs. (12) - (15) with $\chi_{I}$ given by Eqs. (1) -
(9). Our theoretical and experimental results on $C_{q}$ are
summarized in Table II and compared with the Figure 5 (for $q$=2,3)
and Figure 6 (for $q$=4,5), indicating that the KNO scaling is
approximately valid at ISR energies but with clear indication that
it is broken above $\approx$ 100 $GeV$. At the LHC energies from
$\sqrt{s}=7000$ to 14000 $GeV$ the theoretical results predicts
strong violation of the scaling, in qualitative agreement with the
results reported by CMS Collaboration, albeit in pseudorapidity
interval of $|\eta|$ $ < $ 2.4 \cite{CMS2011}.

We recall that the $C_{q}$ moments depends uniformly on the
probabilities and in its calculation the lowest multiplicities are
suppressed and the high multiplicity tail is enhanced. Thus, our
results reflects the influence of the tail of the distributions.

\section{Concluding Remarks}

Applying a phenomenological procedure in which the eikonal is
written as the sum of quark-quark, quark-gluon and gluon-gluon
contributions, we present theoretical predictions for both inelastic
cross section and $C_{q}$ moments of $P_{n}$ for $pp$ interactions.
The comparisons of predictions with a variety of published data
shows good agreement. The imaginary eikonal $\chi_{_{I}}(s,b)$
energy dependence, and hence of the $\sigma_{in}$, has been
completely determined only from elastic fit. The $P_{n}$ values has
been obtained by adopting an approach which enables linkage between
the elastic and inelastic channels thought unitarity condition of
the $S$-matrix. Our mathematical procedure allows studies of the
parton-parton inelastic cross sections, $\sigma_{in\,;\,ij}$,
dependence on $\sqrt{s}$, as well as partonic inelastic overlap
function, $G_{in\,;\,ij}$, where $(i,j)=(q,g)$. Based on the
approximate results on the collision energy dependence of
$G_{in\,;\,ij}$, we have discussed the violation of KNO scaling as a
possible consequence of the manifestation of semihard partons in the
particle production mechanism. At the LHC energies our results
predicts strong violation of the KNO hypothesis. This result is in
agreement with several experimental and theoretical previous studies
developed on the subject. The limited use of the one-channel eikonal
approach to simultaneously describe the total and the elastic cross
sections has been briefly discussed and prediction to the ratio
$\sigma_{el}/\sigma_{tot}$ as function of $\sqrt{s}$ is presented
and compared with the data. Despite some simplifications made in our
procedure, reflecting only approximate results on the collision
energy dependence of the partonic inelastic cross sections and
overlap functions, we believe that the results may serve as guidance
for a theoretical understanding of the cross sections behavior in
terms of the proton components. An interesting result of this
analysis is that we have a clear idea on the approximate behavior of
the parton-parton components as function of the collision energy in
proton interactions.

\begin{acknowledgments}

I am thankful to M.J. Menon and E.G.S. Luna for helpful discussions
and suggestions. I thank to P.V.R.G. Silva, M.J. Menon and D.A.
Fagundes for the permission to use the experimental data compiled in
\cite{Daniel003}. I am also thankful to two anonymous referees for
valuable comments, suggestions and discussions.

\end{acknowledgments}

\newpage

\begin{table}[htb!]
\caption{\label{tabpi} Experimental data with error bar and
theoretical $C_{q}$ values calculated in this work by using the Eqs.
(10) - (15  ). Data points for $P_{n}$ from  \cite{ABC}
\cite{Alexopoulos}.}

\begin{tabular}{ccccc}
\hline $\sqrt{s}$ - $GeV$ & $C_{2}$ & $C_{3}$ & $C_{4}$ & $C_{5}$ \\
\hline
30.4 & 1.29 $\pm$ 0.05 & $1.97 \pm$ 0.09 & $3.45\pm 0.21$ & $ 6.68\pm 0.52$ \\
     & 1.27            & $1.93$          & $3.35        $ & $ 6.48$         \\
\hline
44.5 & 1.28 $\pm$ 0.04 & $1.95 \pm$ 0.07 & $3.40\pm 0.17$ & $ 6.58\pm 0.47$ \\
     & 1.28            & $1.94$          & $3.38        $ & $ 6.55$         \\
\hline
52.6 & 1.29 $\pm$ 0.03 & $1.98 \pm$ 0.06 & $3.48\pm 0.15$ & $ 6.81\pm 0.42$ \\
     & 1.28            & $1.95$          & $3.40        $ & $ 6.63$         \\
\hline
62.2 & 1.29 $\pm$ 0.03 & $1.97 \pm$ 0.06 & $3.40\pm 0.14$ & $ 6.43\pm 0.33$ \\
     & 1.27            & $1.91$          & $3.28        $ & $ 6.24$         \\
\hline
300  & 1.34 $\pm$ 0.02 & $2.21 \pm$ 0.04 & $4.26\pm 0.07$ & $ 9.23\pm 0.17$ \\
     & 1.35            & $2.23$          & $4.27        $ & $ 9.20$         \\
     \hline
546  & 1.41 $\pm$ 0.03 & $2.52 \pm$ 0.05 & $5.31\pm 0.10$ & $ 12.72\pm 0.24$ \\
     & 1.43            & $2.57$          & $5.41        $ & $ 12.88$         \\
     \hline
1000 & 1.41 $\pm$ 0.02 & $2.47 \pm$ 0.05 & $5.11\pm 0.13$ & $ 11.87\pm 0.36$ \\
     & 1.42            & $2.49$          & $5.11        $ & $ 11.74$         \\
     \hline
1800 & 1.47 $\pm$ 0.02 & $2.78 \pm$ 0.03 & $6.23\pm 0.07$ & $ 15.91\pm 0.21$ \\
     & 1.48            & $2.79$          & $6.20        $ & $ 15.63$         \\
     \hline
7000 & 1.58            & $3.19$          & $7.46        $ & $ 19.21$         \\
     \hline
8000 & 1.60            & $3.25$          & $7.69        $ & $ 20.06$         \\
 \hline
13000 & 1.65           & $3.51$          & $8.71        $ & $ 23.88$         \\
 \hline
14000 & 1.66            & $3.60$          & $9.16        $ & $ 25.91$         \\
\hline
\end{tabular}
\end{table}

\begin{figure}
\label{difdad} \vspace{2.0cm}
\begin{center}
\includegraphics[height=.60\textheight]{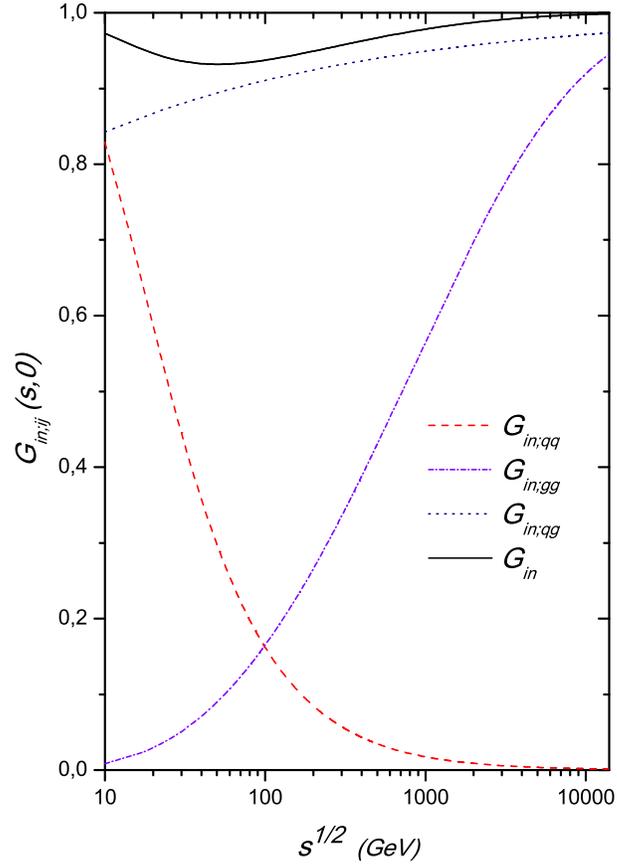}
\caption{The approximate collision energy dependence of the
quark-quark, quark-gluon and gluon-gluon inelastic overlap functions
calculated by using the Eq. (20) at $b \sim $ 0, as explained in the
text. The solid line represents the result for $G_{in}$ obtained
with the Eq. (18).}
\end{center}
\end{figure}

\begin{figure}
\label{difdad} \vspace{2.0cm}
\begin{center}
\includegraphics[height=.60\textheight]{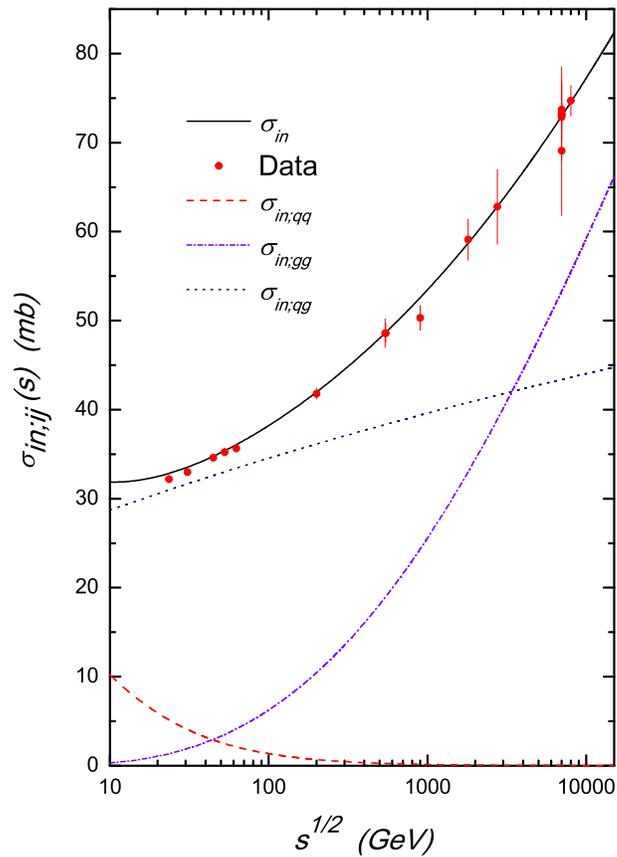}
\caption{Theoretical predictions as a function of $\sqrt{s}$ for
quark-quark, quark-gluon and gluon-gluon inelastic cross sections,
Eq. (22). The solid line represents the total inelastic cross
section, Eq. (21), compared with a variety of data points from
\cite{Aad001} \cite{Abelev001} \cite{Antchev001} \cite{Antchev002}
\cite{Antchev003} \cite{Amaldi} \cite{ABC}.  }
\end{center}
\end{figure}

\begin{figure}
\label{difdad} \vspace{2.0cm}
\begin{center}
\includegraphics[height=.60\textheight]{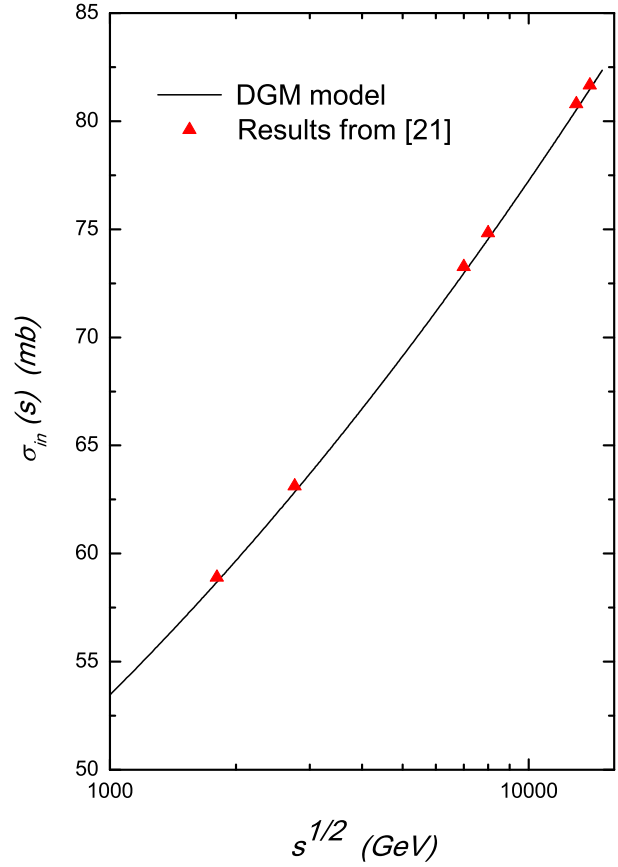}
\caption{Comparison between our calculations for $pp$ inelastic
cross section (solid line) with the predictions from model discussed
in \cite{Erasmo001} at $\sqrt{s}$ of 1800, 2760, 7000, 8000, 13000
and 14000 $GeV$ (triangles).}
\end{center}
\end{figure}

\begin{figure}
\label{difdad} \vspace{2.0cm}
\begin{center}
\includegraphics[height=.60\textheight]{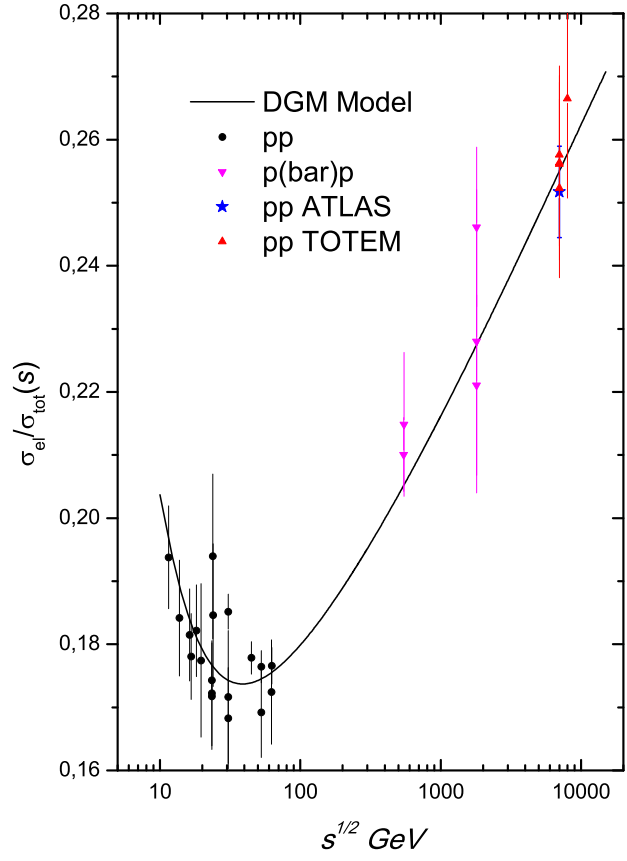}
\caption{DGM model prediction on the ratio $\sigma_{el} /
\sigma_{tot}$ by using the parameter set reproduced from
\cite{BeggioLuna}, Table I. Experimental data are from $pp$
scattering as compiled at \cite{Daniel003}.}
\end{center}
\end{figure}

\begin{figure}
\label{difdad} \vspace{2.0cm}
\begin{center}
\includegraphics[height=.60\textheight]{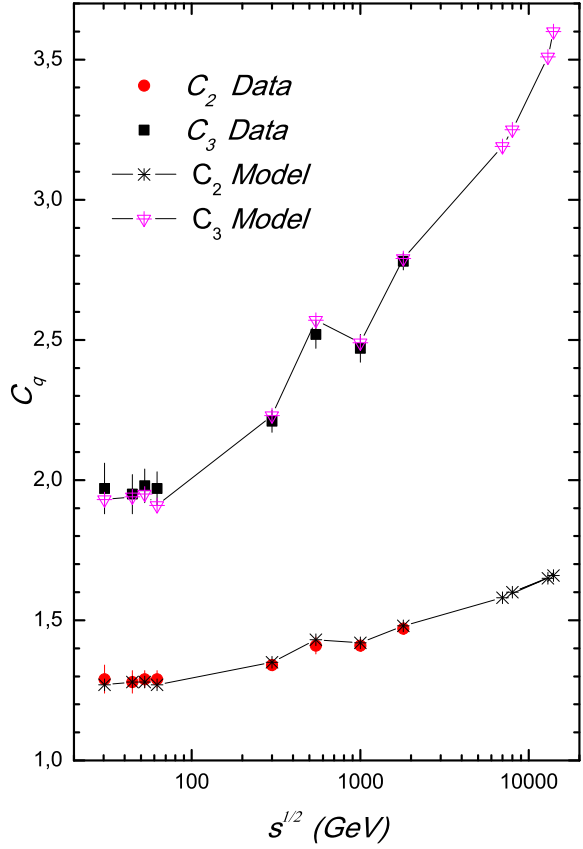}
\caption{Theoretical and experimental $C_{q}$ moments, $q$=2,3,
calculated in this work in full phase space using the Eq. (11).
Theoretical values of $P_{n}$ were obtained applying the Eqs. (12) -
(15) with the imaginary eikonal $\chi_{I}$ given by Eqs. (1) - (9).
Experimental $P_{n}$ values are from \cite{ABC} \cite{Alexopoulos}
and the lines are draw only as a guidance of the theoretical
points.}
\end{center}
\end{figure}

\begin{figure}
\label{difdad} \vspace{2.0cm}
\begin{center}
\includegraphics[height=.60\textheight]{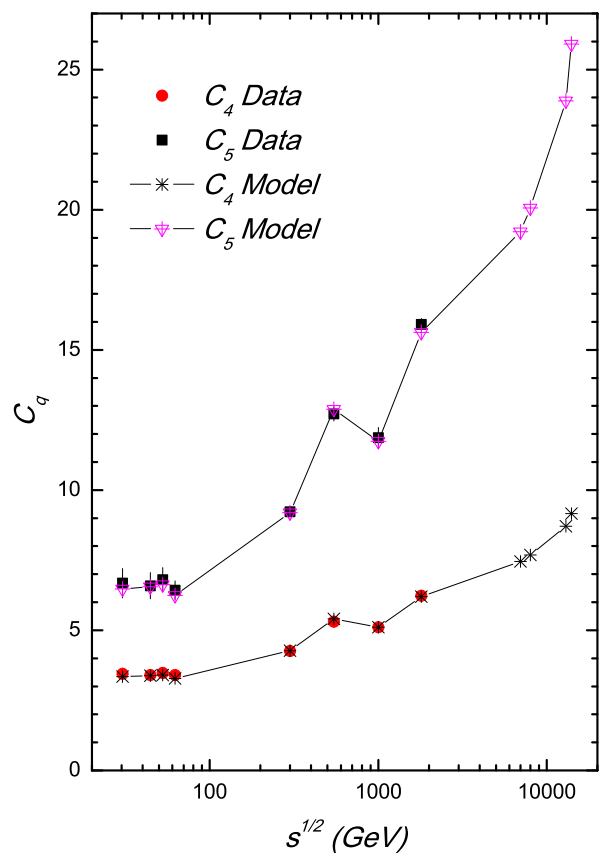}
\caption{Same as Figure 5 but for $q$=4,5.  }
\end{center}
\end{figure}

\end{document}